\documentclass[twocolumn]{aa}  
\usepackage{amsmath,amssymb,amsfonts}
\usepackage{graphicx}
\usepackage{xcolor}
\usepackage{color}
\usepackage{orcidlink}
\usepackage{txfonts}
\usepackage[normalem]{ulem}

\usepackage{hyperref}
\hypersetup{
    colorlinks=true,
    linkcolor=blue,
    citecolor=blue,
    filecolor=magenta,      
    urlcolor=blue,
    pdftitle={The environment of TeV Halo progenitors},
    pdfpagemode=FullScreen,
    }

\begin{document} 
\title{The environment of TeV halo progenitors}

\author{Lioni-Moana Bourguinat\inst{1,2}\thanks{lioni-moana.bourguinat@gssi.it}~\orcidlink{0009-0006-5823-643X}
\and Carmelo Evoli\inst{1,2}~\orcidlink{0000-0002-6023-5253}
\and Pierrick Martin\inst{3}~\orcidlink{0000-0002-7670-6320}
\and Sarah Recchia\inst{4}~\orcidlink{0000-0002-1858-2622}}

\institute{Gran Sasso Science Institute (GSSI), Viale Francesco Crispi 7, 67100 L’Aquila (AQ), Italy
\and
INFN-Laboratori Nazionali del Gran Sasso (LNGS), via G. Acitelli 22, 67100 Assergi (AQ), Italy
\and
IRAP, Université de Toulouse, CNRS, CNES, F-31028 Toulouse, France
\and 
INAF Osservatorio Astrofisico di Arcetri Largo Enrico Fermi, 5, 50125, Firenze, Italy}

\date{Received Month XX, XXXX; accepted Month XX, XXXX}

  \abstract
   {TeV haloes are extended sources of very-high-energy gamma rays found around some middle-aged pulsars. The emission spanning several tens of parsecs suggests an efficient confinement of the ultra-relativistic lepton pairs produced by pulsars in their vicinity. The physical mechanism responsible for this suppressed transport has not yet been identified. In some scenarios, pair confinement may be linked to the medium the pulsars are located in.}
   {We aim at understanding the type of medium pulsars probe over their lifetime.}
   {We developed a model for the environment probed by moving pulsars, from their birth in core-collapse explosions -- where they receive a natal kick -- until their entry into the interstellar medium. The model involves: (i) a Monte-Carlo sampling of the properties of the massive-star progenitors of pulsars; (ii) a calculation of the structure of the surrounding medium shaped by these progenitors for the two cases of isolated stars and star clusters; and (iii) a computation of the evolution of supernova remnants in these parent environments. Ultimately, from a distribution of neutron star kick velocities, we assess the medium in which pulsars are located as a function of time. We first derived the statistical properties of a fully synthetic Galactic population and then applied the model to a selection of known pulsars to assess the likely nature of their environment.}
   {We show that pulsars escape into the interstellar medium at around $300~\mathrm{kyr}$, significantly later than assumed in the literature. Given our assumptions, all known pulsars with a confirmed TeV halo have high probabilities of still being in their parent environment, which suggests that efficient pair confinement is connected to the region influenced by progenitor stars. To test this, we provide the probability that known pulsars still reside in their parent environment for a list of known pulsars.
   }
   {}


\maketitle


\section{Introduction}

Since the discovery of TeV haloes around the Geminga and B0656+14 pulsars by the HAWC experiment in 2017 \citep{HAWC2017}, and around J0622+3749 by LHAASO in 2021 \citep{lhaaso2021}, theoretical efforts have been dedicated to understanding this source class. These very high-energy ($\gtrsim 100~\mathrm{GeV}$) gamma-ray emissions are produced by the inverse-Compton scattering of leptons accelerated in the pulsar wind nebula of the pulsar with the cosmic microwave background and the interstellar photon fields~\citep{dimauro+2019}.
Known haloes originate from middle-aged pulsars ($\sim100~\mathrm{kyr}$) expected to have escaped their parent supernova remnant (SNR) into the interstellar medium (ISM) \citep{HAWC2017, fang+2022, amato+2024}. Surprisingly, the gamma-ray emission hints at a stronger confinement -- with a suppression factor $ \gtrsim 100$ with respect to average conditions in the Galaxy -- of the high energy ($\gtrsim 100~\mathrm{TeV}$) particles around a region of radius $20-30~\mathrm{pc}$ around the pulsars.

The debate regarding the number of detectable TeV haloes has remained active since their discovery.
Several studies expected the number of detected pulsar haloes to grow rapidly \citep{linden+2017, linden+2018}. However, after almost 10 years of searches, the number of candidate TeV haloes grew to just a dozen with most requiring further spectral and morphological studies for confirmation. A high occurrence rate of TeV haloes around middle-aged pulsars also seems disfavoured by the local positron flux and the properties of the known population of nearby pulsars~\citep{martin+2022}.

This discovery has consequences not only in gamma-ray astronomy but also in cosmic-ray physics. For example, before 2017, an effort was made to understand whether pulsars or dark matter could explain the anomalous positron excess in cosmic rays~\citep{yuan+2015, boudaud+2015}.
The detection of TeV haloes led to the idea that the large-scale cosmic-ray diffusion coefficient of the Galaxy -- deduced from secondary and primary cosmic-ray nuclei fluxes, mainly probing the large diffusive halo of the Milky Way -- is not representative of transport conditions in the disc.
As a consequence of this lower diffusion coefficient, cosmic-ray leptons would take more time to diffuse between nearby sources and the Earth, suffer stronger energy losses particularly at the highest energies, which would make nearby pulsars unable to explain the positron excess~\citep{HAWC2017}. Later, more refined studies showed that pulsars could still explain the positron excess if propagation around them is described in a two-zone framework, with suppressed diffusion within tens of parsec of the source -- the halo -- and  normal standard diffusion at larger distances ~\citep{tang+2019,schroer+2023}.

The theoretical efforts to explain the suppressed diffusion can be divided into three categories, depending on whether the suppression is actually necessary or  considered necessary but either linked to the pulsar progenitor or not.
Belonging to the first category, for example, are models of anisotropic transport along the magnetic fields surrounding the pulsar \citep{liu+2019b, delatorreluque+2022}, or models that take into account non purely diffusive transport \citep{recchia+2021}.
Belonging to the second category, are models where the suppressed diffusion is attributed to turbulence self-generated by the lepton pairs emitted by the pulsar \citep{evoli+2018,mukhopadhyay+2022} through the excitation of plasma instabilities.
Belonging to the third category -- the one relevant for the present work -- include, for instance, the work by \citet{fang+2019}, which proposes that pulsars exhibiting a TeV halo remain inside their parent SNRs, whose interiors could be turbulent, and the work by \cite{schroer+2022}, which suggests that the turbulence responsible for the formation of the TeV halo may be self-generated by protons accelerated by the SNR and escaping the system.

The lepton pairs creating TeV haloes are also expected to emit in other wavelengths, such as radio or X-ray through synchrotron emission. Both should in principle allow us to gain new information on the environment of the pulsar, especially its magnetic field strength or structure \citep{liu+2019a}, and the emitting electron and positron population~\citep{khokhriakova+2024}. In the case of Geminga, no X-ray synchrotron halo could be detected, and~\citet{manconi+2024} gave an upper limit on the magnetic field strength around the pulsar.

Despite intense interest in TeV haloes, there remains no clear consensus on the environments that host TeV-halo pulsars -- critically limiting our ability to pinpoint the physical mechanisms behind the observed suppression of cosmic-ray diffusion. This uncertainty is exemplified by Geminga, whose putative SNR remains undetected, and by PSR B0656+14, which appears projected within the Monogem ring, itself a candidate parent SNR. Since the nature and level of turbulence -- and thus particle propagation -- can vary dramatically depending on the local environment, resolving this ambiguity is essential. In this work, we take a decisive step forward: by reconstructing the past trajectory and environment of each pulsar, we directly investigate the medium traversed throughout its life, providing crucial insights into the origin and properties of TeV haloes.

\citet{vanderswaluw+2003} computed the escape time $t_\mathrm{esc}$ at which a pulsar with speed $v_\mathrm{kick}$ reaches the ISM when escaping from a SNR of energy $E_\mathrm{SN}$ propagating in the uniform ISM of density $n_\mathrm{ISM}$ and found
\begin{equation}
    t_\mathrm{esc} = 14 \left(\frac{E_\mathrm{SN}}{10^{51}~\mathrm{erg}}\right)^{1/3}
                        \left(\frac{v_\mathrm{kick}}{1000~\mathrm{km/s}}\right)^{-5/3}
                        \left(\frac{n_\mathrm{ISM}}{1~\mathrm{cm}^{-3}}\right)^{-1/3}~\mathrm{kyr}.
\end{equation}
This leads to the hypothesis that the middle-aged pulsars creating the haloes are all in the ISM.

However, pulsars are born from massive stars that, whether isolated or in groups, blow powerful winds during their lives. The stellar winds carve rarefied, hot, and turbulent bubbles that can extend up to several hundreds of parsecs~\citep{weaver+1977, maclow+1988, vieu+2022}. Then, the supernova explosion following the birth of the pulsar ejects stellar material inside the bubble, which further modifies the surrounding medium. Accounting for this peculiar environment challenges the usual understanding of the environment of pulsars. Stellar bubbles and their interaction with the SNR were never studied in the literature relative to TeV haloes. 

In this paper, we present in Sect.~\ref{sec:model} a model for the environment probed by pulsars following their formation and the impulse they receive in core-collapse supernova explosions of massive-star progenitors. We provide in Sect.~\ref{sec:synthpop} the statistical description of the environment of pulsars as a function of time, for a fully synthetic Galactic population. In Sect.~\ref{sec:knownpsrs} we apply our model to a selection of known pulsars to statistically determine the nature of the medium they are most likely to reside in, and we discuss in more detail a number of individual objects including the current sample of established TeV haloes. Finally, we conclude in Sect. \ref{sec:conclusions}.

\section{Model}\label{sec:model}

\subsection{Pulsar kinematics}\label{sec:Pulsar kinematics}

    Pulsar kick velocities result from asymmetries in the SN explosion happening at the end of the life of the progenitor massive star \citep{Lambiase+2024}. Assuming a uniform and rectilinear trajectory, the pulsar position depends on kick velocity $v_\mathrm{kick}$ and true age $t_\mathrm{age}$ as $r_\mathrm{PSR} = v_\mathrm{kick} t_\mathrm{age}$.
    
    A recent work using parallax and proper-motion measurements of radio pulsar by \citet{igoshev_observed_2020} suggests using single and bimodal Maxwellian distributions for the kick velocities of old and young pulsars, respectively (young meaning a spin-down characteristic age $\tau_\mathrm{c}<3~\mathrm{Myr}$). Since the middle-aged pulsars that we consider in the present work have a characteristic age comprised in the range of $100-300~\mathrm{kyr}$,  we adopted a bimodal Maxwellian distribution $f_{2\mathcal{M}}$:
    \begin{equation}
    f_{2\mathcal{M}}(v) \mathrm{d} v = w \mathcal{M}(v,\sigma_1) \mathrm{d} v + (1-w) \mathcal{M}(v,\sigma_2) \mathrm{d} v \,,
    \end{equation}
    where $v$ is short for $v_\mathrm{kick}$, $\mathcal{M}(v)$ is the Maxwellian distribution, and the optimal parameters are  $w=0.2$, $\sigma_1=56~\mathrm{km/s}$, and $\sigma_2=336~\mathrm{km/s}$\footnote{It is puzzling that a dip in the velocity distribution for young pulsars, at roughly $200~\mathrm{km/s}$, lies exactly where a peak is found in the velocity distribution for all pulsars. \citet{disberg+2025} find it more likely that the low-velocity peak in the bimodal velocity distribution is due to Poisson noise on the nearby pulsars instead of having a physical origin.}.
    
    We assessed the impact of using different kick velocity distributions on a subset of our results by considering the total sample of pulsars in \citet{igoshev_observed_2020} or previous distributions, such as those from \citet{fauchergiguere+2006} and \citet{hobbs+2005}, adopted in the literature.
    
\subsection{Progenitor stars}\label{sec:Progenitor star}

    We assumed that pulsars are born from massive stars of zero-age main-sequence (ZAMS) masses $M_\mathrm{ZAMS}$ between $8~\mathrm{M}_\odot < M_\mathrm{ZAMS} < 20~\mathrm{M}_\odot$ \citep{sukhbold+2016}. Stars in this mass range follow a power-law initial-mass function (IMF) of index $-2.35$ \citep{kroupa+2003}. Stars with ZAMS masses in the interval $2~\mathrm{M}_\odot < M_\mathrm{ZAMS} < 16~\mathrm{M}_\odot$ are considered B stars, while stars with higher masses are O stars. After reaching the end of the main-sequence (MS) phase, massive stars enter the red supergiant (RSG) and Wolf-Rayet (WR) phases. As progenitor properties, we used the results of \citet{seo+2018} and references therein, which are based on stellar evolution calculations by \citet{ekstrom+2012} for non-rotating stars.
    
    The MS and RSG phase durations, $\tau_\mathrm{MS}$ and $\tau_\mathrm{RSG}$, depend on $M_\mathrm{ZAMS}$ and are given in megayears as \citep{zakhozhay2013,seo+2018}
    \begin{align}
        \log \tau_\mathrm{MS} & \approx 9.96 \, - \, 3.32 \log M_\mathrm{ZAMS} + 0.63 \left(\log M_\mathrm{ZAMS}\right)^2 \nonumber\\
        & + 0.19 \left(\log M_\mathrm{ZAMS}\right)^3 - 0.057 \left(\log M_\mathrm{ZAMS}\right)^. ,\\
        \log \tau_\mathrm{RSG} & \approx -2.76 \log M_\mathrm{ZAMS} + 9.38.
    \end{align}
    The fraction of the lifetime spent as RSG decreases strongly with the mass and amounts to 17\% for a $8~\mathrm{M}_\odot$ B star or 7\% for a $20~\mathrm{M}_\odot$ O star.

    We did not account for the WR phase as it only affects stars with masses $\geq25~\mathrm{M}_\odot$.

    The total mass lost $\Delta M$ during each phase is given in $\mathrm{M}_\odot$ by
    \begin{align}
        \log \Delta M_\mathrm{MS}  &\approx 2.36 \log M_\mathrm{ZAMS} - 3.15,\\
        \log \Delta M_\mathrm{RSG} &\approx 0.85 \log M_\mathrm{ZAMS} - 0.24.
    \end{align}
    For the instantaneous wind mass-loss rate $\dot{M}$ (in solar masses per year) and velocity $u_\mathrm{w}$ (in kilometres per second) in each phase, we followed the prescription of \citet{seo+2018}:
    \begin{align}
        \log \dot{M}_\mathrm{MS} &\approx -3.38 (\log M_\mathrm{MS})^2 + 14.59 \log M_\mathrm{MS} - 20.84, \\
        \log \dot{M}_\mathrm{RSG} &\approx 4.16 \log M_\mathrm{RSG} - 10.27, \\
        \log u_\mathrm{w, B} &\approx 0.21 \log M_\mathrm{MS} + 2.85, \\
        \log u_\mathrm{w, O} &\approx 0.08 \log M_\mathrm{MS} + 3.28, \\
        u_\mathrm{w, RSG} &\approx 1.9 \cdot M_\mathrm{RSG}^{0.85}~\mathrm{km/s},
    \end{align}
    where the quantities are given as a function of the mass of the start at the beginning of each phase, so $M_\mathrm{MS} = M_\mathrm{ZAMS}$ and $M_\mathrm{RSG} = M_\mathrm{ZAMS}-\Delta M_\mathrm{MS}$. The wind luminosity $L_\mathrm{w}$ in ergs per second derives from the above quantities as $L_\mathrm{w} = \dot{M} u_\mathrm{w}^2/2$, which yields the following for the MS:
    \begin{align}
        \log L_\mathrm{w, B} \approx -3.38 \left(\log M_\mathrm{MS}\right)^2 + 15.02 \log M_\mathrm{MS} + 20.36, \\
        \log L_\mathrm{w, O} \approx -3.38 \left(\log M_\mathrm{MS}\right)^2 + 14.77 \log M_\mathrm{MS} + 21.21.
    \end{align}
    The formula for $L_\mathrm{w, RSG}$ is not given as it is not directly used. As explained below, the RSG wind properties are used to determine the pressure balance in the bubble of an isolated star, which requires only $u_\mathrm{w, RSG}$ and $\dot{M}_\mathrm{RSG}$.

\subsection{Structure of the surrounding medium}
\label{sec:surrounding medium}

    The majority of O and B stars are formed in star clusters (SCs) and OB associations \citep{wright2020, portegies+2010}. A number of massive stars are found in relative isolation in the Galactic field, either because they were created this way, or because they were ejected from clusters due to gravitational interactions, thus becoming runaway stars. In the following, we handle two different families of progenitors: isolated massive stars and SCs. The latter category gathers all types of groupings irrespective of size or compactness, and it is where most massive stars in our Galaxy are to be found. \citet{zinnecker+2007} state that only $2\%$ of B stars and $10-25\%$ of O stars are found isolated in the Galactic field. Motivated by the subsequent analysis of \citet{carreterocastrillo+2023}, we adopted the high-end value of $25\%$ for runaway O stars. Due to the IMF favouring low-mass stars, the impact of the number of runaway O stars is modest, and most runaways are actually B stars. Combining these probabilities we find that $10\%$ of the massive stars in the mass range we considered are found in isolation, while the rest is found in SCs and OB associations.

    As a consequence, we considered two different types of surrounding medium for the pulsar progenitors: wind-blown bubbles (WBB) and superbubbles (SB), depending on whether the progenitor star is isolated or belongs to a cluster. In both cases, for simplicity, we assumed spherical symmetry for the environment around a star or cluster located at the centre. We therefore neglected effects such as the proper motions of isolated massive stars or the substructures in stellar clusters, both potentially giving rise to asymmetric developments of the environment.

\subsubsection{Wind-blown bubbles around isolated massive stars}\label{sec:Wind-blown bubbles around isolated massive stars}

    The interaction of supersonic stellar winds from isolated massive stars with a typically warm ISM of density $n_\mathrm{ISM}$ generates what we call here a WBB.
    This results in a stratified bubble-like structure \citep{weaver+1977}. In the immediate surroundings of the star, freely expanding wind of speed $u_\mathrm{w}$ ends in a wind termination shock (WTS) at radius $r_\mathrm{w}$. After the WTS, one finds a bubble of hot rarefied plasma with density $n_\mathrm{b}$ and temperature $T_\mathrm{b}$ made of shocked wind and evaporated interstellar material that extends up to radius $r_\mathrm{b}$ and occupies most of the volume of the structure. The bubble is bounded by a thin and dense shell of density $n_\mathrm{shell}$ found at radius $r_\mathrm{shell}$, which consists of all the ISM material that was swept by the bubble during its expansion.
    
    At the end of the MS phase, massive stars enter the RSG phase for $\gtrsim 100~\mathrm{kyr}$ and lose most of their mass in powerful and dense but slow winds. These slow winds and the short duration of the RSG phase are such that they do not impact the extent of the bubble, but they set a new pressure balance in its innermost regions. We therefore assumed that the size of the bubble at the time of the supernova is set from the MS phase only, while the pressure balance is determined from the RSG wind.
    
    The WTS radius $r_\mathrm{w}$ was obtained by equating the ram pressure of the wind during the RSG phase and the thermal pressure of the bubble interior at the end of the MS phase:
    \begin{align}
        r_\mathrm{w} = \left(\frac{\dot{M}_\mathrm{RSG}u_\mathrm{RSG}}{4\pi \mu n_\mathrm{b}\mathrm{k_B} T_\mathrm{b}}\right)^{1/2}.
    \end{align}
    In general, a star of a certain mass in our range exhibits more powerful winds in the RSG than the MS phase, so the WTS positions we obtained (of the order of 10\,pc) are several times the WTS positions obtained from MS winds (of order 1\,pc) as in a number of other similar studies \citep[e.g.][]{dwarkadas2005}.
    The bubble density $n_\mathrm{b}$ and the bubble temperature $T_\mathrm{b}$ were taken from \citet{castor+1975} and \citet{weaver+1977}, who model the evaporation of the shell matter into the hot bubble:
    \begin{align}
        n_\mathrm{b} =  8\times10^{-3}~{\rm cm}^{-3}
                &\left(\frac{\zeta_\mathrm{L}}{0.22}\right)^{6/35} 
                \left(\frac{L_{\rm w, MS}}{10^{36}~{\rm erg/s}} \right)^{6/35}\\
                & \times \left(\frac{n_{\rm ISM}}{{\rm cm}^{-3}} \right)^{19/35}
                \left( \frac{\tau_{\rm w}}{\rm Myr} \right)^{-22/35}, \nonumber\\
        T_\mathrm{b} = \ 1.46 \times 10^6~{\rm K}
                &\left(\frac{\zeta_\mathrm{L}}{0.22}\right)^{8/35} 
                \left(\frac{L_{\rm w, MS}}{10^{36}~{\rm erg/s}} \right)^{8/35}\\
                & \times \left(\frac{n_{\rm ISM}}{{\rm cm}^{-3}} \right)^{2/35}
                \left( \frac{\tau_{\rm w}}{\rm Myr} \right)^{-6/35}.\nonumber
    \end{align}
    The wind time $\tau_\mathrm{w}$ contributing to the expansion of the bubble is defined as $\tau_\mathrm{w} = \tau_\mathrm{MS} + \tau_\mathrm{RSG}$.
    
    We added a corrective factor $\zeta_\mathrm{L}$ on the luminosity to account for the effect of radiative losses, possibly in relation to mixing and turbulence at the bubble-shell interface. Radiative losses are not included in \citet{weaver+1977} for the evolution of WBBs, leading to bubble sizes bigger than observed \citep{yadav+2017}. Lacking a value specific to WBBs and expecting the behaviour to be similar to SBs, we adopted the value of $\zeta_\mathrm{L}=22~\%$ proposed in \citet{harer+2023} and based on the observation of SBs.
    
    The bubble radius $r_\mathrm{b}$ is given in parsecs from \citet{weaver+1977}:
    \begin{align}
        r_\mathrm{b} = 21~\mathrm{pc} \left(\frac{\zeta_\mathrm{L}}{0.22}\right)^{1/5} \left(\frac{L_\mathrm{w, MS}}{10^{36}~\mathrm{erg/s}}\right)^{1/5} \left(\frac{\tau_\mathrm{w,MS}}{1~\mathrm{Myr}}\right)^{3/5} \left(\frac{n_\mathrm{ISM}}{1~\mathrm{cm}^{-3}}\right)^{-1/5}.
    \end{align}
    The wind density profile is obtained by $\rho_\mathrm{w} = \dot{M}/4\pi u_\mathrm{w} r^2$. We recall that the mass $\rho$ and number $n$ densities, used interchangeably throughout this paper, are linked by $\rho = \mu m_\mathrm{p} n$ with $\mu$ the molecular fraction taken at 1.4 and $m_\mathrm{p}$ the proton mass.
    
    After the bubble, the shell contains all the ISM matter that was in the volume of the bubble at the star's birth and that was swept up by the expanding bubble. The shell width is assumed to be $\Delta r_\mathrm{shell}=1~\mathrm{pc}$, and the volume of the shell lies between $r_\mathrm{b}$ and $r_\mathrm{shell} = r_\mathrm{b} + \Delta r_\mathrm{shell}$:
    \begin{align}
        n_\mathrm{shell} = n_\mathrm{ISM} \frac{r_\mathrm{b}^3}{r_\mathrm{shell}^3-r_\mathrm{b}^3}.
    \end{align}
    The shell and ISM are both considered a warm medium, so from \citet{recchia+2022}, we take the corresponding temperatures $T_\mathrm{shell}=T_\mathrm{ISM}=8000~\mathrm{K}$. We provide in Table \ref{tab:parameters_WBB} the properties of the WBB at the time of supernova for two stellar progenitors of initial mass $M_\mathrm{ZAMS}=8~\mathrm{M}_\odot$ and $M_\mathrm{ZAMS}=20~\mathrm{M}_\odot$.

    \begin{table}[]
        \begin{tabular}{ccc}
        \hline \hline
        Progenitor Mass {[}M$_\odot${]}       & 8                  & 20                 \\ \hline 
        MS time {[}Myr{]}                     & 37.8               & 9.6                \\
        Ejecta mass {[}M$_\odot${]}           & 3.1                & 10.4              \\
        RSG mass loss rate {[}M$_\odot$/yr{]} & $2.9\times10^{-7}$ & $1.2\times10^{-5}$ \\
        RSG wind speed {[}km/s{]}             & 11.0               & 23.4              \\
        MS wind luminosity {[}erg/s{]}        & $1.5\times10^{31}$ & $5.1\times10^{34}$ \\
        RSG Wind radius {[}pc{]}              & 16.4               & 17.2              \\
        Bubble density {[}cm$^{-3}${]}        & $5.3\times10^{-5}$ & $5.0\times10^{-4}$ \\
        Bubble temperature {[}K{]}            & $6.2\times10^{4}$  & $5.0\times10^{5}$  \\
        Bubble radius {[}pc{]}                & 20.1               & 45.4               \\
        Shell density {[}cm$^{-3}${]}         & 4.5                & 10.6             \\
        Shell radius {[}pc{]}                 & 21.1               & 46.4               \\
        Shell Mass {[}M$_\odot${]}            & 841                & 12793 \\              \hline
        \end{tabular}
        \caption{Properties of the stellar progenitors and associated wind-bubble environments at the time of supernova for the extreme values of the initial mass range considered in our work.}
        \label{tab:parameters_WBB}
        
    \end{table}
    
\subsubsection{Superbubbles around massive SCs}\label{sec:Superbubbles around massive star clusters}

    Superbubbles can be described in the same theoretical framework as WBBs \citep{weaver+1977, harer+2023}, replacing the wind properties of a single star by the collective mechanical power of all stars of the cluster. As a result, SBs extend over hundreds of parsecs instead of tens of parsecs for WBBs.

    In this work, we neglected any structure in the stellar cluster or its spatial extent and assumed that SBs are the result of a mechanical outflow emanating from a central point. For each SB, we first sampled the total cluster mass $M_\mathrm{cl}$ between $10^3~\mathrm{M}_\odot \lesssim M_\mathrm{cl} \lesssim 10^5~\mathrm{M}_\odot$ from the following young massive SC IMF \citep{portegies+2010},
    \begin{align}
        \frac{\mathrm{d}N}{\mathrm{d}M_\mathrm{cl}} \propto M_\mathrm{cl}^{-2} \exp(-M_\mathrm{cl}/M_\mathrm{cl, cut}),
    \end{align}
    with the cutoff mass $M_\mathrm{cl, cut}=2\times10^5~\mathrm{M}_\odot$ for spiral galaxies. We then populated the SC with stars of mass $M_\mathrm{ZAMS}$ taken from a Salpeter IMF with index $\alpha=2.35$,
    \begin{align}
        \frac{\mathrm{d}N}{\mathrm{d}M_\mathrm{ZAMS}} \propto M_\mathrm{ZAMS}^{-\alpha} \exp(-M_\mathrm{\mathrm{ZAMS}, cut}/M_\mathrm{ZAMS}),
    \end{align}
    for masses between $0.1~\mathrm{M}_\odot<M_\mathrm{ZAMS}<150~\mathrm{M}_\odot$, and adopted a cutoff mass of $M_\mathrm{\mathrm{ZAMS}, cut}=0.35~\mathrm{M}_\odot$ \citep{Larson1998}. We sampled from this distribution until the cluster's mass was reached.

    The SC properties needed for modelling the SB are its mechanical luminosity $L_\mathrm{w,SC}$, its collective wind mass loss rate $\dot{M}_\mathrm{SC}$, and speed $u_\mathrm{w,SC}$. The first two quantities are simple sums of the luminosity $L_\mathrm{w}$ and mass loss rate $\dot{M}$ of individual massive stars $i$:
    \begin{align}
        L_\mathrm{w,SC} &= \sum_i L_{\mathrm{w}, i},   \\
        \dot{M}_\mathrm{SC} &= \sum_i \dot{M}_i.
    \end{align}
    In contrast, the SC collective wind speed $u_\mathrm{w,SC}$ is linked to the wind speeds $u_\mathrm{w}$ and mass loss rates of individual stars using momentum conservation as in \citet{menchiari+2024}:
    \begin{align}
        u_\mathrm{w,SC} = \frac{\sum_i u_{\mathrm{w},i}\dot{M}_i}{\dot{M}_\mathrm{SC}}.
    \end{align}
    In 1D, all the stellar winds contribute fully to the wind luminosity, while in reality some energy is lost through wind interactions. The losses remain marginal when considering the bubble size. This was shown in \citet{vieu+2024} where the bubble still evolves by following the \citet{weaver+1977} model, even if they show an absence of WTS around the OB association.
    As stellar wind properties of individual stars, we adopted a different approach than that followed in Sect.~\ref{sec:Wind-blown bubbles around isolated massive stars} for the WBBs, for the sake of simplicity (because the superposition of different winds from a high number of stars at different evolutionary stages is a complex problem). First, we considered O and B stars with masses $>2~\mathrm{M}_\odot$ (and not just for $8~\mathrm{M}_\odot < M_\mathrm{ZAMS} < 20~\mathrm{M}_\odot$, which was motivated by the ability to form a pulsar), as this broader mass range yields a better account of the total energetics of the cluster. Second, we restricted the wind properties to those of the MS stage, which we picked from \citet{seo+2018}. Last, we assumed that the collective mechanical power of the SC remains constant and equal to its initial value, when all its member stars are in the MS stage. Thus, we did not account for any change in luminosity when stars of the cluster leave the MS and enter RSG or WR stages, and we also missed the decrease in luminosity when the massive stars exploded. Conversely, we did not consider the energy input from SN explosions. Nevertheless, this contribution was assessed in \citet{vieu+2022}, and the authors showed that the mechanical power of a cluster is roughly constant over the first tens of millions of years, which suggests that our assumption of a constant luminosity is reasonable. A more realistic description of a SB including these effects would have to rely on hydrodynamical simulations of the time-dependent outflows from an ageing star population, but this complex modelling goes beyond the scope of our statistical work.

    Finally, \citet{parizot+2004} found that SCs are born in giant molecular clouds (GMCs) with a typical gas number density $n_\mathrm{GMC} \sim 100~\mathrm{cm}^{-3}$. We took this value for the density of the ambient medium surrounding the SB, thus setting $n_\mathrm{ISM} = n_\mathrm{GMC}$.
    
    Out of all the stars able to produce pulsars (we recall $8~\mathrm{M}_\odot<M_\mathrm{ZAMS}<20~\mathrm{M}_\odot$) in the cluster, we randomly picked one and computed its MS time. Contrary to the WBB case, when the selected massive star that produces a pulsar dies, the rest of the cluster can still act as an engine for the SB; therefore, we evolved the system after the SN of the selected star until the pulsar reached the shell and escaped.

\subsection{Supernova remnant evolution}\label{sec:SNR evolution}

    At the end of its life, the progenitor massive star explodes as a core-collapse SN, which leads to the formation of a pulsar with a natal kick velocity and the ejection of stellar material at supersonic speeds. A SNR results from the expansion and interaction of the latter in the aforementioned parent environments. We here describe how the SNR dynamics was computed in our model.
    
    The SNR parameters are the SN energy, for which we used the fiducial value of $E_\mathrm{SN}=10^{51}~\mathrm{erg}$, and the ejecta mass $m_\mathrm{ej}$. It could be possible to add a dispersion on $E_\mathrm{SN}$, but observationally this parameter is degenerate with the medium density, resulting in loose constraints on the parameter. Therefore, we kept it at its commonly assumed value. The ejecta mass was computed as the progenitor mass minus the mass lost in winds and the remaining pulsar mass $m_\mathrm{PSR}=1.4~\mathrm{M}_\odot$:
    \begin{align}
        m_\mathrm{ej} = M_\mathrm{ZAMS} - (\Delta M_\mathrm{MS} + \Delta M_\mathrm{RSG} + m_\mathrm{PSR}).
    \end{align}

    There are no comprehensive analytical models for the evolution of a SNR inside a stratified environment similar to a bubble that also incorporates the radiative phase and merger. We used the thin-shell approximation for the evolution of the SNR blast wave radius and speed in the stratified environment to which we added a criterion for the merger with the medium surrounding the SNR.

    The model is based on the method given in Appendix A of \citet{ptuskin+2005} for the computation of the SNR position $r$ as a function of time $t$. We recall the equations used here, beginning with the equation for the SNR mass $M(r) = M_\mathrm{ej} + M_\mathrm{sw}(r)$, with $M_\mathrm{sw}(r)$ representing the swept-up mass until the shock position $r$:
    \begin{align}\label{eq:mass_integral}
        M(r) = M_\mathrm{ej} + 4\pi \int_0^r \mathrm{d}r' r'^2 \rho(r'),
    \end{align}
    where $\rho(r)$ corresponds to the density profile.

    The SNR mass enters in the shock velocity $u_\mathrm{SNR}$ profile expression:
    \begin{align}\label{eq:shock_speed}
        u_\mathrm{SNR}(r) = \frac{\gamma+1}{2} \left[\frac{2 \alpha E_\mathrm{SN}}{M^2(r) r^\alpha} \underbrace{\int_0^r \mathrm{d}r' r'^{\alpha-1}M(r')}_{I(r)}\right]^{1/2},
    \end{align}
    with $\gamma=5/3$ being the adiabatic index for a monatomic gas in the medium and $\alpha=6(\gamma-1)/(\gamma+1)$.

    Finally, the evolution of the shock position as a function of time is given by
    \begin{align}\label{eq:shock_position}
        t(r) = \int_0^r \frac{\mathrm{d}r'}{u_\mathrm{SNR}(r')}.
    \end{align}

    The SNR speed profile can be solved analytically by assuming a density profile. We recall that the density profile we use follows:
    \begin{equation}
        \rho(r) =
        \begin{cases}
            \rho_{\rm w}(r)  & r < r_{\rm w},               \\
            \rho_{\rm b}     & r_{\rm w} < r < r_{\rm b},   \\
            \rho_{\rm shell} & r_{\rm b} < r < r_{\rm shell}, \\
            \rho_{\rm ISM}   & r > r_{\rm shell}.
        \end{cases}
    \end{equation}
    
    We solved Eq.~\ref{eq:shock_speed} analytically in this density profile and Eq.~\ref{eq:shock_position} numerically to finally find the SNR's age as a function of its position. When and if the SNR reaches the bubble shell, it sweeps the mass emitted by the progenitor star in the form of MS and RSG winds, as well as some interstellar matter evaporated from the shell as a result of thermal conduction \citep{weaver+1977}. Yet, the shell made up of swept-up ISM material is much heavier than the SNR and effectively acts as a solid wall. As a result, the SNR collides and very rapidly loses most of its energy radiatively, as discussed both theoretically and observationally in \citet{vink2020}, to merge with the shell. In our model framework, the SNR therefore has an impact on the bubble interior but never reaches the ISM. In case the SNR does not encounter the shell, it can merge with its surrounding medium at the position $r_\mathrm{merge}$ when it stops being a strong shock:
    \begin{align}
        u_\mathrm{SNR}(r_\mathrm{merge})=\beta c_\mathrm{s}(r_\mathrm{merge}),
    \end{align}
    where $\beta=2$ \citep{cioffi+1988} is an arbitrary parameter that gives the minimal shock strength and $c_\mathrm{s}(r)=\sqrt{\gamma k_\mathrm{B}T(r)/\mu m_\mathrm{p}}$ is the speed of sound in the medium.

    In this analytical framework, the radiative losses of the SNR were not considered. This modelling can only reproduce correctly the ejecta dominated and Sedov-Taylor phases of the SNR. This means that the position of the SNR as a function of time is always lightly overestimated in these calculations, especially at the later times.

\section{Synthetic population of pulsars}
\label{sec:synthpop}

In this section, we present the results of our model on the type of environment crossed by a moving pulsar as a function of time, up to pulsar ages of a few megayears. In the following, we use the concept of 'parent environment' to refer to the structured medium that surrounds a pulsar at birth. This immediate-vicinity environment is determined by the nature of the progenitor star (isolated or in a cluster), as well as by the properties of the SNR resulting from the core-collapse explosion, whose expansion modifies the surrounding medium. We first introduce the properties of this parent environment in relation to our model parameters, before presenting the statistical implications for the medium experienced by pulsars along their trajectory.

\subsection{Properties of the parent environment}

\begin{figure}
        \centering
        \includegraphics[width=0.45\textwidth]{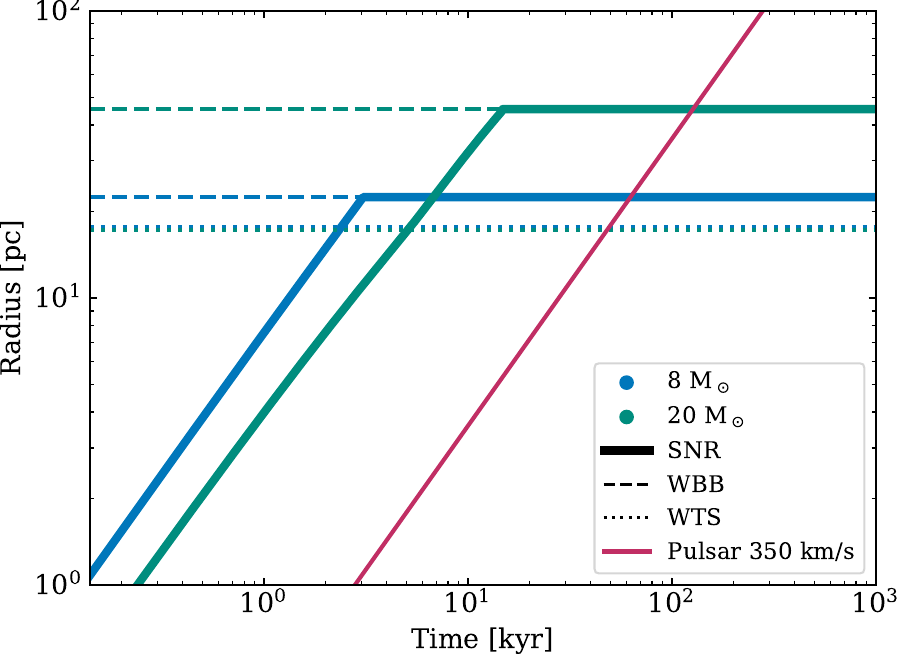}
        \caption{Evolution of the SNR as a function of time in the WBB scenario, for two values of the stellar progenitor initial mass. For each case, we overplotted the positions of the wind-termination shock (dotted lines, on top of each other, see Tab. \ref{tab:parameters_WBB}) and outer bubble radius (dashed lines) at the time of explosion. The purple line indicates the trajectory of a pulsar with kick velocity $350~\mathrm{km/s}$.}\label{im:SNR_WBB}
    \end{figure}

    Figure \ref{im:SNR_WBB} shows the evolution of SNRs for the two extreme values of progenitor star mass in the range we considered, $8~\mathrm{M}_\odot$ and $20~\mathrm{M}_\odot$, along with the corresponding wind-bubble structure at the time of supernova. In both cases, the expansion of the SNR is halted by its encounter with the bubble shell, at $3~\mathrm{kyr}$ and 15$~\mathrm{kyr}$, respectively. This is due to the mass of the bounding shell being much higher than that of the expanding remnant, by more than one order of magnitude (see Table \ref{tab:parameters_WBB}). In the WBB case, a typical pulsar with a kick velocity of $350~\mathrm{km/s}$ escapes from the bubble shell and enters the ISM at $64~\mathrm{kyr}$ and $130~\mathrm{kyr}$, respectively. 
    
\begin{figure}
        \centering
        \includegraphics[width=0.45\textwidth]{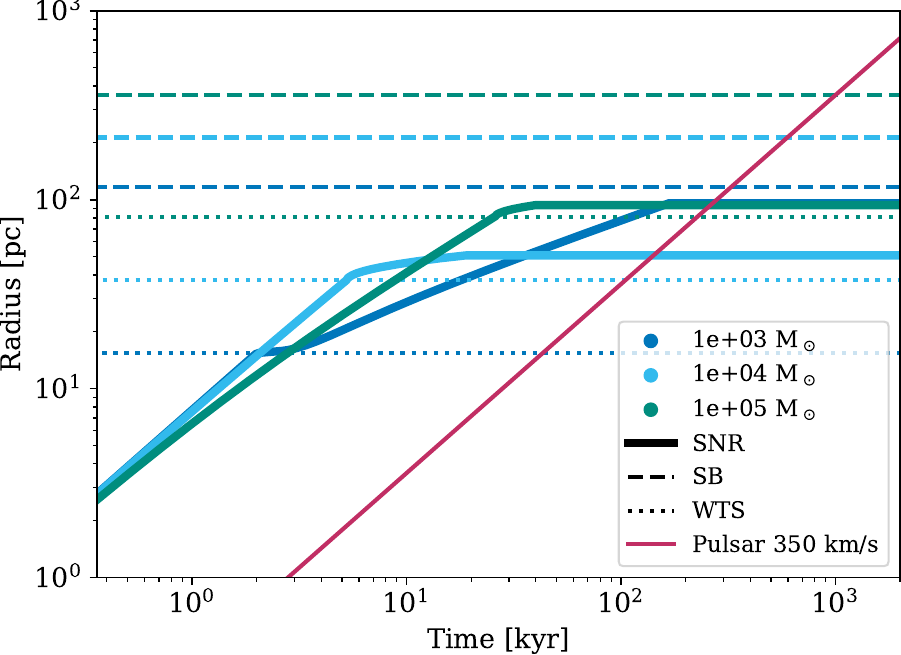}
        \caption{Evolution of the SNR as a function of time in the SB scenario, for a stellar progenitor of initial mass $8~\mathrm{M}_\odot$ and three different masses of the stellar cluster it belongs to. Graphical elements are similar to Fig. \ref{im:SNR_WBB}. The SNR curves turn flat when the remnant merges with the cavity interior and disappears.}\label{im:SNR_SB}
    \end{figure}

    Figure \ref{im:SNR_SB} shows the evolution of SNRs for a progenitor star of mass $8~\mathrm{M}_\odot$ in a SB, for three different values of the cluster mass, $10^3~\mathrm{M}_\odot$, $10^4~\mathrm{M}_\odot$, and $10^5~\mathrm{M}_\odot$. As expected, the higher the cluster mass, the larger the SB radius because of the more powerful collective wind. In these examples, rather large SB are obtained because the chosen progenitor star has a long lifetime, during which the bubble could expand to large distances. Since the parameters of the explosion are the same, the trajectory changes because of the density structure inside the bubble cavity, which is primarily determined by the cluster luminosity and age, and by the exterior density. We see in all three cases that the SNR evolves from an ejecta-dominated phase to a Sedov-Taylor phase, before merging inside the SB without even reaching the bubble shell. When this happens, a typical pulsar with a kick velocity of $350~\mathrm{km/s}$ remains inside the remnant. 
    
    We notice that the SB radius in the most massive case of $10^5~\mathrm{M}_\odot$ is $320~\mathrm{pc}$, which is larger than the typical scale height of the gas disc of our Galaxy \citep{langer+2014}. A proper treatment would require us to model the evolution of the SB in at least 2D with a gas density gradient, as well as a proper modelling of the stellar population in the cluster but this is beyond the scope of our study.
    In addition, because of the cluster IMF, this case is statistically rare, so we simply neglect these considerations.

    In the adopted model framework, the WBB and SB shells are the relevant boundaries marking the entry into the ISM, and not the forward shock of the SNR (as would be the case when not taking into account a modified circumstellar medium). Before a pulsar reaches the ISM, it can be either inside its parent SNR, the WBB, or the SB (owing to their small widths, we neglect the short time spent in the bounding shells of the bubbles). When the pulsar resides in the bubble medium, it is interesting to distinguish two different setups: one in which the SNR is still alive and expanding, and one in which it has dissolved in the WBB or SB interior. Yet, in our Monte-Carlo simulations and given the adopted pulsar kick velocities, the latter case appears to be more common.
    Given the range of progenitor masses, in the WBB case, the SNR always reaches the bubble shell before the pulsar, namely the pulsar can either be found inside the SNR or in the ISM.
    In the SB case, the prospects are more interesting. A pulsar can escape a still-expanding SNR and be found in the SB interior for a cluster mass $10^3~\mathrm{M}_\odot$ -- the most favoured by the cluster IMF -- but only if it has a kick velocity typically exceeding $800~\mathrm{km/s}$ and a massive progenitor of $20~\mathrm{M}_\odot$. However, the most probable situation for the SB case is that the SNR merges with the bubble cavity while the pulsar was still moving within. Observationally, such a setup could have relevant consequences, particularly in the context of pulsar haloes, since it implies the possibility of finding a young or middle-aged pulsar in what is likely a turbulent medium (the interior of the bubble), even though the parent SNR can no longer be detected. Geminga may very well be in such a situation.
    
\subsection{Statistical description of a pulsar environment}

    For each family of progenitor stars described above (isolated or in cluster), we generated a population of 10000 pulsars in their parent environments using a Monte-Carlo method. This was achieved by random sampling the initial star mass for WBB systems, the cluster and the progenitor star masses for SB systems, and the pulsar kick velocities in both cases.
    
\begin{figure}
        \centering
        \includegraphics[width=0.45\textwidth]{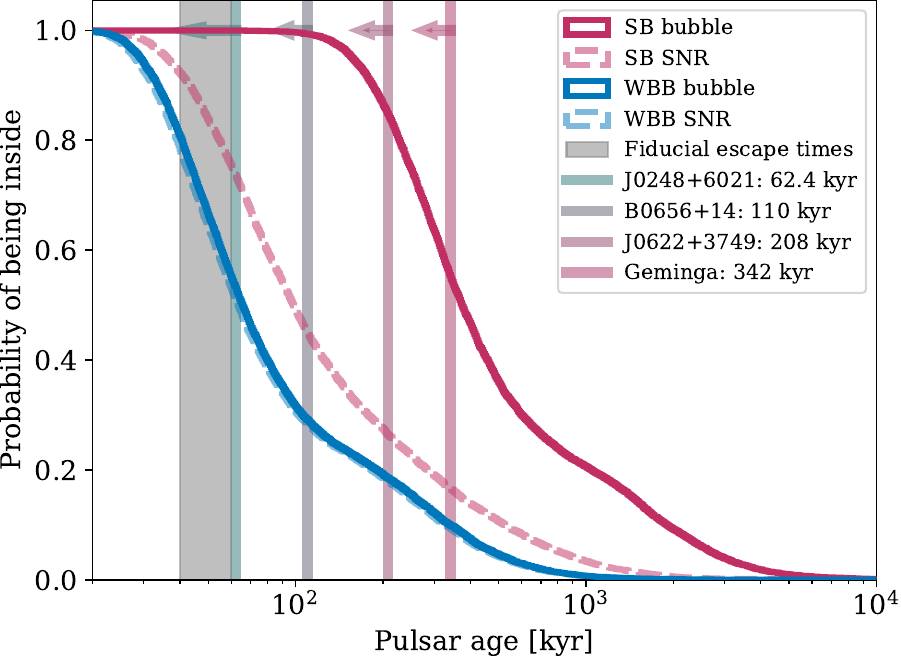}
        \caption{Probability for a pulsar to be found inside its parent remnant or bubble environment as a function of time. The red and blue lines correspond to the SB and WBB cases, respectively. The grey-shaded band corresponds to values typically found in the literature for the time at which a pulsar exits into the ISM, thus showing that they are significantly underestimated. The four vertical lines correspond to the characteristic ages of the confirmed TeV haloes.}
        \label{im:escape_times_galaxy_detail}
    \end{figure}

    From these mock populations, we computed the probability as a function of age that a pulsar is located in a particular medium among those two: the interior of the SNR and the interior of the WBB or SB. A null probability means that the pulsar has escaped into the ISM. We plotted the corresponding results in Fig. \ref{im:escape_times_galaxy_detail}, using as pulsar kick velocity distribution that of \citet{igoshev_observed_2020} since it is based on the most complete pulsar sample. For comparison, we overplotted as vertical coloured lines the characteristic ages of the first four pulsar haloes discovered by HAWC (B0656+14 and Geminga) and LHAASO (J0622+3749 and J0248+6021), and as a shaded band the classical values of $40-60~\mathrm{kyr}$ routinely found in the literature for the time at which a pulsar enters the ISM \citep[e.g.][]{gaensler+2006,amato+2024}.
    
    Our population synthesis actually predicts that pulsars will enter the ISM somewhat later than usually assumed: at an age of 60\,kyr, more than $52\%$ of pulsars in the WBB case and all pulsars in the SB case still reside in their parent environment. As expected due to the smaller size of the structure, pulsars remain inside WBBs for a shorter amount of time compared to SBs. In the latter case, the transition from pulsars being mostly in to pulsar being mostly out occurs at an age of $200-500$\,kyr.
    
\begin{figure}
        \centering
        \includegraphics[width=0.45\textwidth]{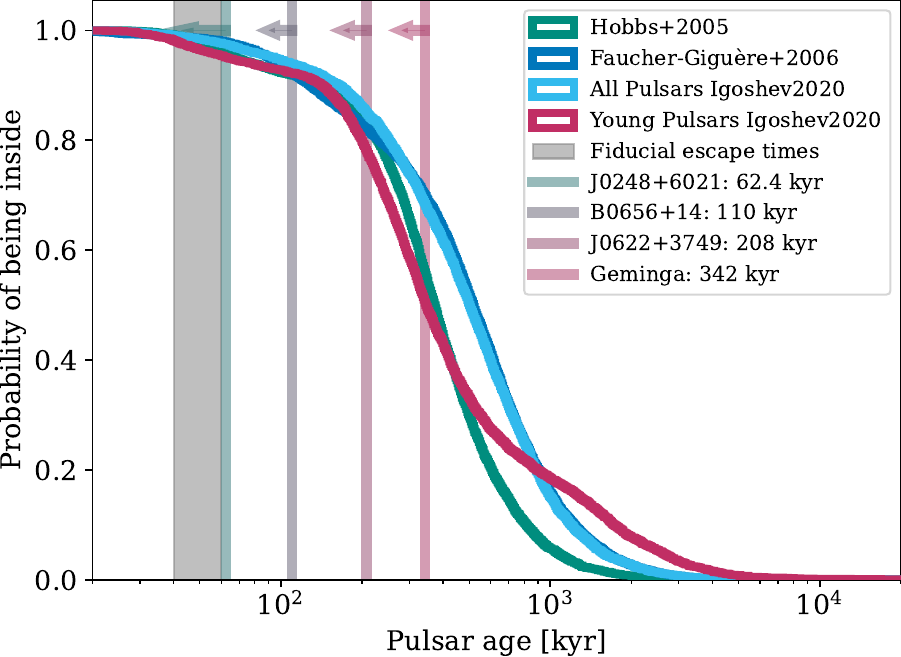}
        \caption{Probability for a pulsar to be found inside its parent environment as a function of time, for a Galactic population made of a proper mix of isolated and cluster massive-star progenitors (see text). The different curves correspond to different assumptions on the pulsar kick velocity distribution. Other graphical elements are similar to Fig. \ref{im:escape_times_galaxy_detail}.}
        \label{im:escape_times_galaxy}
    \end{figure}

    Both the WBB and SB cases can be combined, with the proper weighing for the global proportion of isolated versus clustered massive stars (see Sect. \ref{sec:surrounding medium}), to make predictions for a Galactic population. This is illustrated in Fig. \ref{im:escape_times_galaxy}, which resembles strongly the SB case in Fig. \ref{im:escape_times_galaxy_detail} since most massive stars live in clusters.  Overall, it can be seen that before $30~\mathrm{kyr}$, almost all pulsars are inside their parent environment, while all escaped after $6~\mathrm{Myr}$. Most pulsars escape into the ISM at $\sim100-1000$\,kyr, with half of the Galactic population being out at $300~\mathrm{kyr}$. This is precisely the age range of the so-called middle-aged pulsars presumably involved in TeV haloes. 
    
    From our model and population synthesis, there is therefore no clear-cut interpretation of TeV haloes in terms of the medium a pulsar lies in, since middle-aged pulsars have comparable chances of being inside their (possibly turbulent) parent environment or in the (a priori more quiescent) ISM.
    Conversely, if we assume that the parent environment, provides enough turbulence for efficient trapping of ultrarelativistic electron-positron pairs, and that this is the main formation channel for TeV haloes, our model predicts a strongly decreasing occurrence rate of TeV haloes as pulsars get older than 300\,kyr. At the characteristic age of Geminga, half of the pulsars would then be able to form haloes. Our model also provides the complementary information that about half of these halo-forming pulsars would still be located inside their parent SNR (see Fig. \ref{im:escape_times_galaxy_detail}). This is qualitatively consistent with the proposal put forward in \citet{martin+2022} that pulsar haloes may not be so common around middle-aged pulsars.
    
    As can be see in Fig. \ref{im:SNR_SB} in the SB case, the SNR can merge before reaching the bubble shell. A pulsar can be observed without a SNR surrounding it but still being inside the SB. Increasing the required shock strength before merger from $\beta=2$ to $\beta=5$ makes the SNR merge with the interior of the bubble marginally earlier, slightly increasing the time the pulsar spends inside the SB without probing the downstream of a SNR.
    
    Finally, in Fig. \ref{im:escape_times_galaxy}, we illustrate the impact of the kick velocity distribution adopted in the Monte-Carlo sampling. We see a clear impact on the position of the transition -- whether predominantly in or out of the parent environment -- and, to a lesser extent, on its shape. As a consequence, the quantitative statements made above or below should be treated with some caution.
    
\section{Application to a selection of known pulsars}
\label{sec:knownpsrs}

\subsection{Sample construction}
\label{sec:Sample Construction}

    We applied our model to a sample of real pulsars in order to derive constraints on their environments, in a statistical sense. From the ATNF catalogue \citep{manchester+2005}, we extracted a subset of pulsar halo candidates with properties that are relatively similar to Geminga, the prototypical pulsar halo. 
    
    We considered pulsars with luminosity to distance ratio $\dot{E}/d \geq 10^{33}~\mathrm{erg/s/kpc}$ and characteristic age $40~\mathrm{kyr}<\tau_{\mathrm{age}}<1~\mathrm{Myr}$. The first criterion aims at jointly maximising the power and angular size of the halo emission, while the second is meant to discard that are too young systems and are more likely to be classical PWNe. We added a criterion to restrict our sample to relatively nearby objects. For a TeV halo to be detected as an extended source with current experiments, an upper limit needs to be set on the distance of the system.  Assuming a typical halo size of $r_\mathrm{halo}\sim30~\mathrm{pc}$ and considering that it would be detected as extended with HAWC and LHAASO if spanning at least three point-spread functions of $\alpha_\mathrm{PSF}\sim0.3~\mathrm{deg}$, we derived an upper limit on the halo visibility at distance $d_\mathrm{max}=r_\mathrm{halo}/\tan(\alpha_\mathrm{PSF})=2~\mathrm{kpc}$. We complemented the sample with the four well-established pulsar TeV haloes (J0633+1746, B0656+14, J0622+3749 and J0248+6021) and three other objects not fulfilling the above criteria but mentioned in the literature as good halo candidates, mainly on the basis of gamma-ray observations \citep[B0540+23, J0633+0632, and J0631+1036; see][]{celli+2024,khokhriakova+2024}. Finally, our set of real pulsars was divided into those having estimated transverse velocity and those without, as the former category allows us to make a finer treatment and to derive richer constraints on their environments. The full sample is listed in Table \ref{table:pulsar_properties}.
    
    \subsection{Model predictions}
    
    \begin{figure}
        \centering
        \includegraphics[width=0.45\textwidth]{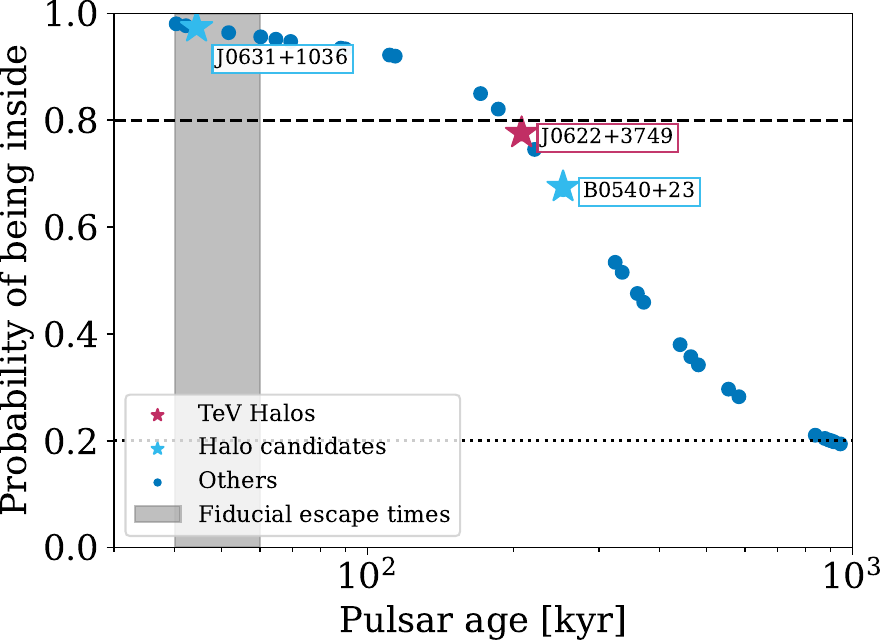}
        \caption{Probability for a selection of 30 known pulsars within $2~\mathrm{kpc}$ to be found inside their parent environments, as a function of their characteristic ages. This is based on a Galactic population model made of a proper mix of isolated and cluster massive-star progenitors (see text), and a kick velocity distribution from \citet{igoshev_observed_2020}. Dashed and dotted lines mark the $80\%$ and $20\%$ probabilities, respectively.}
        \label{im:prob_novkick}
    \end{figure}
    
    \begin{figure*}
        \centering
        \includegraphics[width=\textwidth]{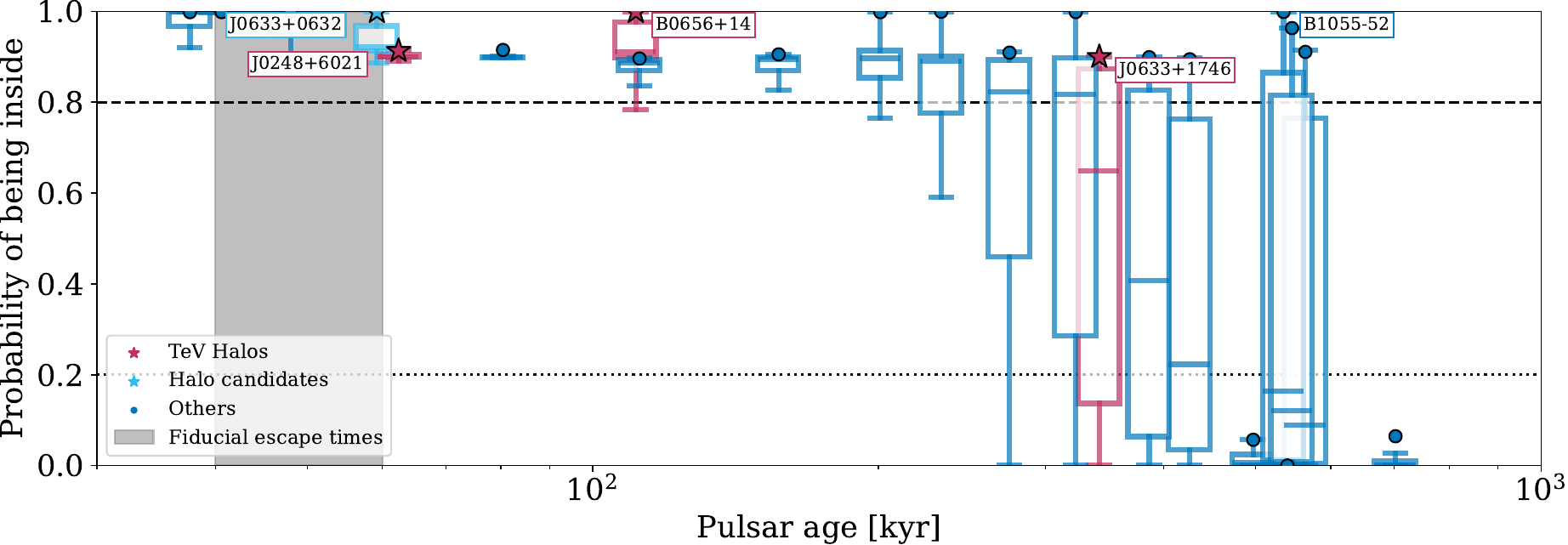}
        \caption{Probability for a selection of 22 known pulsars within $2~\mathrm{kpc}$ to be found inside their parent environments. Pulsars in this sample have proper motion information $v_\mathrm{t}$, which enters as a further constraint on their actual situation (see text). Coloured symbols with black edges indicate the probability for $v_\mathrm{kick} = v_\mathrm{t}$. Boxes indicate the 68\% containment interval over possible kick velocities $v_\mathrm{kick} \geq v_\mathrm{t}$, while whiskers extending above and below the boxes show the 95\% containment interval. The horizontal segment inside each box is the median probability.}
        \label{im:prob_vkick}
    \end{figure*}
    
    Before discussing the predictions of our model for the selected known pulsars, we recall that the characteristic age of a pulsar should actually be understood as an upper limit to the true pulsar age. In cases where a more accurate estimate of the age of the pulsar can be obtained by other methods, it appears that the characteristic age can markedly overestimate the true age, by up to one order of magnitude \citep[][see also the cases of PSR J0908-4913 and PSR J0538+2817 in Sect. \ref{sec:Discussion of individual pulsars}]{suzuki+2021}. In the absence of a reliable estimate for the age of the pulsars in our sample, we use their characteristic age in the following discussion; nonetheless, the caveat above should be kept in mind.
    
    From the age of selected pulsars without velocity information, we can plot Fig. \ref{im:prob_novkick} based on the combined WBB and SB case for a Galactic population. There, 13 (one) out of 30 pulsars in that sub-sample have a probability higher than $80\%$ (below $20\%$) of being found inside their parent environment. Such an analysis can be useful to prioritise targets in searches for pulsar haloes, depending on the confinement scenario to be tested. The two halo candidates in this sample (J0631+1036, and B0540+23) and the established TeV halo (J0622+3749) have probabilities $\gtrsim 65$\% of still residing in their parent environment. In the case of J0631+1036, the probability reaches $\gtrsim95$\%, making it a particularly good target for testing whether the parent environment provides the conditions for pair confinement.
    
    The sample of pulsars with known transverse velocities is smaller but allows us to extract more information since transverse velocity is a crucial parameters in our model. We still consider for this analysis the combined WBB and SB case for a Galactic population.
    First, we recall that transverse velocity $v_\mathrm{t}$ is the projection on the sky of the kick velocity $v_\mathrm{kick}$ of the pulsar such that $v_\mathrm{kick} = v_\mathrm{t}/\cos(\theta)$ with $\theta$ the angle between the two velocities. A measured $v_\mathrm{t}$ therefore results from any value $v_\mathrm{kick} \geq v_\mathrm{t}$, each implying a different kinematics of the pulsar with respect to the parent environment. The probability of each $v_\mathrm{kick}$ is given by the distribution of natal kick velocities from \citet{igoshev_observed_2020}, plus a weighing $\propto \cos(\theta)$ to account for the fact that a uniform distribution of kick orientations in space statistically favours velocity vectors close to the plane of the sky (the allowed solid angle is larger for a given $d\theta$ element).
    
    For each pulsar with a given characteristic age and transverse velocity, we randomly sampled a high number of possible values of $v_\mathrm{kick} \geq v_\mathrm{t}$ (based on the distributions of kick velocities and orientations mentioned above), and we computed for each $v_\mathrm{kick}$ the probability for the pulsar to be located in its parent environment. We then defined ranges for the most frequently encountered probabilities together with confidence intervals. The result is displayed in Fig. \ref{im:prob_vkick} in the form of a box plot. For each object in our sub-sample, the following graphical elements are provided: coloured symbols with black edges indicate the probability for the smallest possible velocity $v_\mathrm{kick} = v_\mathrm{t}$; boxes indicate the 68\% containment interval, while whiskers extending above and below the boxes show the 95\% containment interval; last, the horizontal segment inside each box is the median probability.
    
    Since $v_\mathrm{t}$ is the smallest possible kick velocity and is sometimes much smaller than the actual $v_\mathrm{kick}$, using it in our model leads to a high probability that most objects, even quite old pulsars, remain inside their parent environment. Applying the above calculation corrects for this bias and avoids overestimating this probability.
    As a result and in agreement with previous discussions, pulsars younger than $\tau_\mathrm{age}\lesssim300~\mathrm{kyr}$ still have a high probability of being inside their parent environment, while there is a transition from inside to outside between $300~\mathrm{kyr}\lesssim\tau_\mathrm{age}\lesssim600~\mathrm{kyr}$. 
    All pulsars associated with a TeV halo or a halo candidate have high probabilities of still residing inside their parent environment according to our model.

\subsection{Discussion of individual pulsars}
\label{sec:Discussion of individual pulsars}

    We provide here a more detailed discussion of a number of specific pulsars and their most likely environment in view of our model's predictions. 
    We focussed on the pulsars in our sample associated with a PWN, because those were the targets of in-depth studies, which provide more information about the context of the system (for instance the association with a SNR, the evolutionary state or system age, or the system viewing geometry). We cross-matched our selection of pulsars with the list of PWNe provided in \citet{olmi2023}, keeping the distinction made between normal and fast-moving pulsars (translated into the PWN and SPWN identifiers in Table \ref{table:pulsar_properties}). Our selection includes nine pulsars with PWNe. We aimed at classifying these nine pulsars into the following two categories : those that can be located with respect to their putative parent remnant, and those that exhibit typical signatures of a propagation in the ISM (in principle, a pulsar can belong to the two categories, but this did not happen here).
    
    The first category is well-defined, while the second category is less straightforward. The association of a pulsar with a bow-shock pulsar wind nebula (BSPWN) could in principle indicate that it is travelling in the ISM. Indeed, prototypical BSPWNe are expected to develop when pulsars propagate in a cold and mostly neutral ISM with a relatively low sound speed, and they are characterised by a double shock structure and optical line emissions generated through collisional excitation and charge exchange~\citep{olmi2023}. Yet, the identification to a textbook BSPWN condition is far from simple. There are many circumstances that can give rise to a BSPWN appearance in X-rays or radio (high pulsar proper motion and/or recent interaction with the reverse shock of the remnant), while the typical signatures of a bow shock are not easily detectable or interpretable (complex geometry, low-surface-brightness extended emission, and lack of optical emission lines owing to upstream ionisation by the pulsar itself). Each case requires a dedicated investigation.
    
    Six pulsars in our sample are listed in \citet{kargaltsev+2017} and \citet{olmi2023} as fast-moving pulsars with morphological features reminiscent of BSPWNe, and three are associated with more classical PWNe. Out of these, three display strong evidence for a connection with a SNR (J1809-2332, J0538+2817, and J0908-4913), to which we added B0656+14 (for reasons detailed below), while two seem to be travelling in the ISM (J0742-2822 and J1741-2054). We discard J0357+3205, J1740+1000, and J0358+5413 since none of the available information on these pulsars can be used in the present context (the first two are high-latitude objects that may be not representative of the pulsar population). Finally, we provide a specific discussion of J0633+1746, the Geminga pulsar and the prototypical halo.

\subsubsection{Pulsars likely associated with a SNR}

    \paragraph{PSR J1809-2332}
    The pulsar is associated with SNR G7.5-1.7, detected as a partial radio shell \citep{roberts+2008}, and powers the Taz PWN. It lies in projection within the remnant, at about half a radius from the centre. The proper motion measured from a decade of X-ray observations points back towards the centre of the SNR and suggests a kinematic age of $48~\mathrm{kyr}$ \citep[][about 30\% less than the characteristic spin-down time]{vanetten+2012}. The picture is fully consistent with the prediction of our model, which places the pulsar inside its parent environment with a high probability of $95-100$\%.
    
    \paragraph{J0538+2817}
    The pulsar is associated with SNR Sim 147 and lies well within it in projection, even accounting for the estimated total kick velocity \citep[see][and references therein]{khabibullin2024a}. The angular displacement from the geometrical centre of the remnant suggests a kinematic age of $38~\mathrm{kyr}$, which is much smaller than the characteristic spin-down time of $618~\mathrm{kyr}$. We therefore adopted the former value in lieu of the latter. Interestingly, the properties of the SNR suggest that the explosion of the progenitor star took place in a wind-blown cavity \citep{khabibullin2024b}, similar to our WBB case or to our SB case for a low-mass cluster. For such a small age, our model unambiguously places the pulsar inside its parent environment with a high probability.
    
    \paragraph{PSR J0908-4913/B0906-49}
    \citet{johnston+2021} updated the distance of the pulsar from $1~\mathrm{kpc}$ in the ATNF catalogue to $3~\mathrm{kpc}$, which increases the transverse velocity from $240~\mathrm{km/s}$ to $670~\mathrm{km/s}$. A SNR is associated with PSR J0908-4913, and the pulsar lies in projection on its edge, on the verge of escaping it. Our model predicts a $\sim88$\% probability of being inside its parent environment. Yet, there are caveats in this comparison. First, the association with the SNR suggests a kinematical age of $12.7~\mathrm{kyr}$, significantly lower than the characteristic age of $110~\mathrm{kyr}$ of the pulsar (that we used in Fig. \ref{im:prob_vkick}). Reducing the age of the system to such a lower value would increase the probability of it being inside the parent environment up to almost 100\%. On the other hand, the physical size of the putative SNR is $10~\mathrm{pc}$ at a distance of $3~\mathrm{kpc}$, which is smaller than the values we obtain at around $10~\mathrm{kyr}$, even in the WBB case. This would tend to pull the probability to lower values. A dedicated modelling of this object with different parameters (lower explosion energy or wind mass loss rate for instance) would be required to get a more solid conclusion. But overall the model predicts probabilities of residing in the parent environment higher than 88\%, which is consistent with observations of a pulsar in the downstream of the forward shock of its SNR.
    
    \paragraph{PSR B0656+14}
    This pulsar lies in projection inside the Monogem ring, which has been suggested to be its parent SNR \citep{knies+2018}. With its characteristic age of $110~\mathrm{kyr}$, PSR B0656+14 has a median probability of $91\%$ of still residing in its parent environment in our model, consistent with this picture. The fact that a TeV halo was detected around B0656+14 strongly suggests that the confinement of particles on the scale of several tens of parsecs has to do with the conditions inside the remnant or bubble. We notice that the radius of the SNR and ISM density are respectively $r=87~\mathrm{pc}$ and $n_\mathrm{H}=4\times10^{-3}$\,H\,cm$^{-3}$, for a distance of $300~\mathrm{pc}$ and using a Sedov-Taylor evolutionary solution. Such low densities can be expected from SB cavities, and it is therefore possible that the progenitor of B0656+14 exploded in a SB environment, which enabled the SNR to grow to such a large size without being stopped by the SB shell lying at $\gtrsim100$\,pc (see Fig. \ref{im:SNR_SB}).

\subsubsection{Pulsars likely travelling in the ISM}
    
    \paragraph{J0742-2822/B0740-28}
    This pulsar sports a clear optical bow shock \citep{jones+2002} and, as such, could be considered to be in the predominantly neutral ISM, away from its parent environment. With $\tau_\mathrm{c} = 157~\mathrm{kyr}$, PSR J0742-2822 is in the age range where the transition between mostly in to mostly out of the parent environment just starts to occurs. Accordingly, our model predicts a probability distribution from 85 to 90\% (95\% confidence interval). The observed stand-off distance favours small inclination of the kick velocity vector with respect to the plane of the sky \citep{jones+2002}, which would in turn favour high probabilities of residing in the parent environment. On the other hand, the morphology of the optical nebula suggests that it recently left a low-density medium ($n_\mathrm{H} \lesssim 3\times10^{-3}\,\mathrm{H}\,\mathrm{cm}^{-3}$) to enter a denser medium ($n_\mathrm{H} \simeq 3\times10^{-1}\,\mathrm{H}\,\mathrm{cm}^{-3}$), which can be interpreted as a transition from the interior of a bubble to the bounding shell of ambient ISM.
    
    \paragraph{PSR J1741–2054}
    The pulsar displays a clear bow-shock nebula in optical lines \citep{mignani+2016,romani+2010} and UV \citep{abramkin+2025}. This strongly suggests the pulsar propagates in the mostly neutral ISM, which is weakly consistent with our model prediction of an approximately median 60\% probability that it resides outside its parent environment. Both \citet{camilo+2009} and \citet{abramkin+2025} noticed that the properties of J1741-2054 are very close to that of the 'Three Musketeers' (Geminga, B0656+14, and B1055-52). Since TeV haloes were discovered around the first two, and that B1055-52 is in the blind spot of HAWC, J1741-2054 appears as a promising target to search for a TeV halo and make a comparison with the most established instances of the phenomenon. If a halo were discovered around J1741-2054, our model would suggest pair confinement is linked to some properties of the ISM rather than be due to the parent environment.

\subsubsection{The case of Geminga}

    With a characteristic age of $342~\mathrm{kyr}$ and a transverse speed of $150~\mathrm{km/s}$, PSR J0633+1746 has travelled at least $\sim 53~\mathrm{pc}$ since its birth. Because of this large distance, it is commonly assumed that Geminga is in the ISM.
    
    However, our model statistically places the TeV haloes and halo candidates in our sample inside their parent environment, including Geminga. Since Geminga is usually taken as the prototypical pulsar halo, we provide three additional elements that further support the idea that it is not probing the ISM.
    
    First, as summarised in \citet{amato+2024}, there are many hints that Geminga is still in a hot environment. The non-detection of H$_\alpha$ lines in the near vicinity of the pulsar suggests that the medium Geminga traverses is hot and highly ionised. The ambient medium density inferred from Geminga's X-ray jets is quite low, $n_\mathrm{H} \lesssim 10^{-2}$\,H\,cm$^{-3}$, which is typical of a bubble interior. This seems consistent with the fact that Geminga is located close to the Gemini H$_\alpha$ ring, a bubble that hosts an OB association. Using the formalism from \citet{weaver+1977} and comparing with Suzaku observations of the H$_\alpha$ ring, \citet{knies+2018} shows that the Gemini ring can be seen as a bubble blown by at least two stars of an OB association found inside. \citet{fang+2019} noticed that Geminga and the brightest part of its halo are found in projection inside the Gemini H$_\alpha$ ring.
    
    Second, the morphology of X-ray jets around Geminga suggests that the kick velocity of the pulsar is nearly perpendicular to the line of sight (hence in the plane of the sky), especially when jets are interpreted as bent polar outflows \citep{posselt+2017}. In Fig. \ref{im:prob_vkick}, this implies that the actual position of Geminga is closer to the star symbol, which implies a probability of residing in its parent environment above 90\%.
    
\begin{figure}
        \centering
        \includegraphics[width=0.45\textwidth]{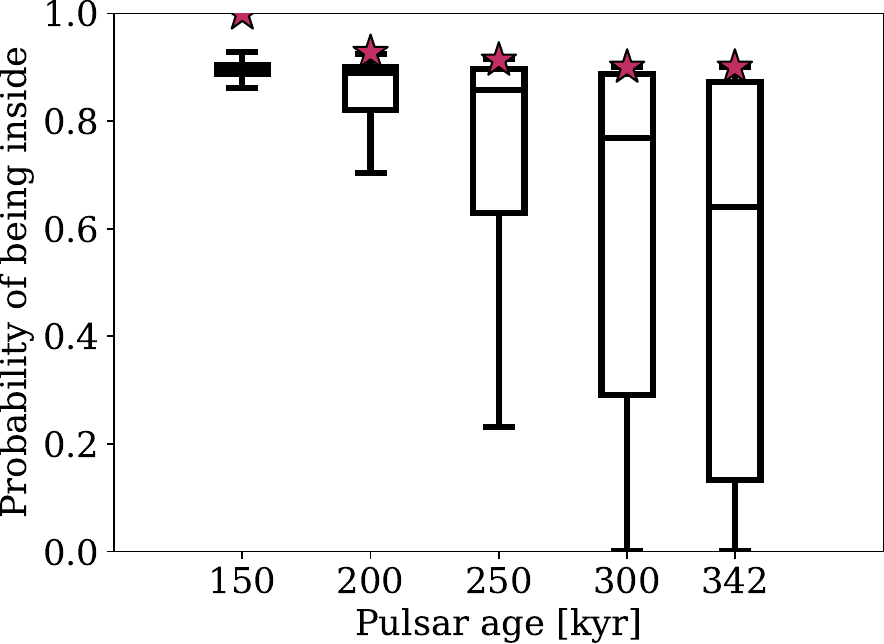}
        \caption{Probability that Geminga resides in its parent environment for five different assumptions on its age. For each age, a box plot such as those in Fig. \ref{im:prob_vkick} is displayed.}
        \label{im:prob_geminga}
    \end{figure}

    Third and last, the discussion so far is based on the characteristic age of Geminga. As mentioned earlier, the characteristic age may well overestimate the true age by a factor of a few (a caveat valid for all objects in our sample). Since Geminga is exactly at the age range where most pulsars transitions from inside to outside of their parent environment, any decrease in age has significant consequences. This is illustrated in Fig. \ref{im:prob_geminga}, where we display box plots similar to Fig. \ref{im:prob_vkick} for a variety of age assumptions for Geminga between $150-342~\mathrm{kyr}$. We see that if Geminga's true age is roughly half of its characteristic age, its median probability of being inside its parent environment reaches $\gtrsim 90\%$.
    
\section{Conclusions}\label{sec:conclusions}

    In this work, we aimed at understanding the type of medium crossed by pulsars over their lifetime in order to constrain the physical mechanism leading to strong confinement of electron-positron pairs and ultimately to the formation of gamma-ray haloes observed at greater than or approximately equal to TeV photon energies around some pulsars.
    We developed a statistical model for the environment probed by pulsars over the first few megayears of their life, under two different scenarios for the circumstellar conditions: pulsars resulting from either isolated massive star progenitors surrounded by WBBs at the time of a supernova or massive stars belonging to clusters that eventually explode inside SBs. We also produced a mixed scenario from a weighted combination of these two cases to describe the population of our Galaxy.
    
    Using a Monte-Carlo method to sample over progenitor properties and pulsar kick velocities, we computed the statistical distribution of the times at which pulsars escape their parent SNR first, and later their parent WBB or SB interior, to eventually enter into the undisturbed ISM. We applied this model to a fully synthetic population of pulsars and to a sample of known pulsars that either have the right properties to produce detectable haloes or were claimed to be halo candidates on the basis of gamma-ray observations. Our main conclusions are the following:
    \begin{enumerate}
        \item In the vast majority of cases, the parent remnant disappears before the pulsar can escape it. In WBBs, the SNR evolution is halted by its collision with the much more massive bubble shell, while in the SBs, SNRs merge with the hot interior.
        \item Pulsars escape into the ISM much later than commonly assumed. Only half of a Galactic population of synthetic pulsars have left their parent environments at $300~\mathrm{kyr}$, compared to the typical escape times of $40-60~\mathrm{kyr}$ frequently used in the literature.
        \item In our model, middle-aged pulsars with ages $100-300$\,kyr, those presumably involved in TeV haloes, have higher chances of being inside their parent environment. While this does not favour any explanation for the origin of haloes in terms of the medium they develop into, this has implications on the occurrence rate of TeV haloes: if the parent environment, stellar-wind bubble or SB, provides the turbulence for efficient trapping of ultrarelativistic electron-positron pairs, our model predicts a strongly decreasing occurrence rate of TeV haloes as pulsars get older than 300\,kyr. At the characteristic age of Geminga, a half of pulsars would then be able to form haloes.
        \item From a sample of seven known pulsars with a confirmed or suspected halo, all of them have a significant probability of being inside their parent environment. Geminga has the highest median probability in the sample (only $35\%$) of being in the ISM at face value. Yet, including additional information on the kick velocity orientation and considering the uncertainty of the pulsar age increase the probability that Geminga resides in its parent environment to above $90$\%, consistent with the hot and ionised properties inferred for the medium around Geminga. 
    \end{enumerate}
    
    We provide the probability of residing inside the parent environment for a list of 53 known pulsars. Searching for extended gamma-ray emission around them with HAWC, LHAASO, and, in the future, the upcoming SWGO experiment could help us connect the occurrence of the pulsar halo phenomenon with the environment, at least from a statistical point of view. Together with complementary studies of the ISM or circumstellar medium (for instance searching for hot ionised gas in X-rays, or for neutral gas in optical lines), this should contribute to our knowledge of the mechanisms underlying the formation of TeV haloes.
    
    Beyond these considerations, our model's prediction that pulsars escape their parent environment much later than commonly assumed will have implications for their contribution to the flux of cosmic-ray positrons and electrons directly and locally measured. All else being equal, a delayed escape likely reduces the flux at Earth, as a smaller fraction of the pairs produced by pulsars will be effectively injected in the ISM.
    Accounting for the AMS-02 positron spectrum thus implies either higher pair production efficiency, contributions from a complementary source class, or different cosmic-ray transport assumptions near and/or far from the sources.

\begin{acknowledgements}
This work has made use of NASA's Astrophysics Data System Bibliographic Services.
Lioni-Moana Bourguinat acknowledges the hospitality and support of IRAP, where part of this work was carried out.
Pierrick Martin acknowledges financial support by ANR through the GAMALO project under reference ANR-19-CE31-0014.
The work of Sarah Recchia is funded by the European Union – "NextGenerationEU" RRF M4C2 1.1 under grant PRIN-MUR 2022TJW4EJ.
The work of Carmelo Evoli has been partially funded by the European Union – "NextGenerationEU", through PRIN-MUR 2022TJW4EJ and by the European Union – NextGenerationEU under the MUR National Innovation Ecosystem grant ECS00000041 – VITALITY/ASTRA – CUP D13C21000430001.
\end{acknowledgements}

\bibliography{Bibliography/biblioPulsarHalo}

\begin{appendix}
\onecolumn
\section{Properties of the individual pulsars studied}

This table compiles the properties of all the individual pulsars that fit the conditions given in Sec.~\ref{sec:Sample Construction}. The probability in the second column represents the probability of pulsars without proper motion information being found inside the parent environment, and the median probability (with the $1\sigma$ interval) for pulsars with proper motion information, in both cases corresponding to the combined WBB+SB scenario for a Galactic population. In the last column, we provide the identification as TeV halo or halo candidate, and as PWN for a normal and fast-moving pulsar (PWN and SPWN, respectively). The data shown come from the ATNF catalogue~\citep{manchester+2005} unless a different reference is provided: (1)~\citet{khabibullin2024b}; (2)~\citet{vanetten+2012}; (3)~\citet{danilenko+2019}; (4)~\citet{johnston+2021}; (5)~\citet{pletsch+2012}.

\small

\begin{longtable}{cccccccc}

\caption{Properties of the known pulsars selected for comparison to our model.}\\
\label{table:pulsar_properties} \\
\hline\hline
Pulsar name & Probability {[}\%{]} & $1\sigma$ interval {[}\%{]}              & $\tau_\mathrm{c}$ {[}kyr{]} & $v_\mathrm{kick}$ {[}km/s{]} & $\dot{E}$ {[}erg/s{]} & $d$ {[}kpc{]} & Association    \\ \hline
\endfirsthead
\caption{continued.}\\
\hline
Pulsar name & Probability {[}\%{]} & $1\sigma$ interval {[}\%{]}              & $\tau_\mathrm{c}$ {[}kyr{]} & $v_\mathrm{kick}$ {[}km/s{]} & $\dot{E}$ {[}erg/s{]} & $d$ {[}kpc{]} & Association    \\ \hline
\endhead
\hline
\endfoot
B2334+61    & 99.87                & {[}99.80; 99.90{]}                       & 40.6                        & 50                           & $6.30\times 10^{34}$  & 0.7           & -              \\
J0538+2817  & 99.86                & {[}96.70; 99.90{]}                       & 37.6$^{(1)}$                & 357                          & $4.90\times 10^{34}$  & 1.3           & PWN            \\
J1809-2332  & 99.85                & {[}94.83; 99.90{]}                       & 67.6$^{(2)}$                & 112                          & $4.30\times 10^{35}$  & 0.88          & PWN            \\
J1044-5737  & 97.79                & -                                        & 40.3                        & -                            & $8.00\times 10^{35}$  & 1.895         & -              \\
J0940-5428  & 97.48                & -                                        & 42.2                        & -                            & $1.90\times 10^{36}$  & 0.377         & -              \\
J0631+1036  & 97.24                & -                                        & 44.4                        & -                            & $1.70\times 10^{35}$  & 2.104         & Candidate Halo \\
J0554+3107  & 96.39                & -                                        & 51.7                        & -                            & $5.60\times 10^{34}$  & 2             & -              \\
J1429-5911  & 95.43                & -                                        & 60.2                        & -                            & $7.70\times 10^{35}$  & 1.955         & -              \\
J1459-6053  & 94.98                & -                                        & 64.7                        & -                            & $9.10\times 10^{35}$  & 1.84          & -              \\
J1954+2836  & 94.63                & -                                        & 69.4                        & -                            & $1.00\times 10^{36}$  & 1.96          & -              \\
B1916+14    & 93.51                & -                                        & 88.1                        & -                            & $5.10\times 10^{33}$  & 1.304         & -              \\
J1028-5819  & 93.39                & -                                        & 90                          & -                            & $8.30\times 10^{35}$  & 1.423         & -              \\
J1732-3131  & 92.41                & -                                        & 111                         & -                            & $1.50\times 10^{35}$  & 0.641         & -              \\
J1740+1000  & 92.27                & -                                        & 114                         & -                            & $2.30\times 10^{35}$  & 1.227         & PWN            \\
J0633+0632  & 92.09                & {[}91.31; 97{]}                          & 59.2$^{(3)}$                & 338                          & $1.20\times 10^{35}$  & 1.355$^{(3)}$ & Candidate Halo \\
B0656+14    & 91.16                & {[}90; 97.10{]}                          & 111                         & 60                           & $3.80\times 10^{34}$  & 0.286         & TeV Halo       \\
J0248+6021  & 90.00                & {[}90; 90.73{]}                          & 62.4                        & 644                          & $2.10\times 10^{35}$  & 2             & TeV Halo       \\
B1727-47    & 90.00                & {[}89.91; 90.03{]}                       & 80.4                        & 501                          & $1.10\times 10^{34}$  & 0.7           & -              \\
J2032+4127  & 89.62                & {[}84.76; 91.29{]}                       & 201                         & 19                           & $1.50\times 10^{35}$  & 1.33          & -              \\
B0740-28    & 89.54                & {[}87.29; 90{]}                          & 157                         & 278                          & $1.40\times 10^{35}$  & 2             & SPWN           \\
B1822-09    & 89.19                & {[}78.25; 90.28{]}                       & 233                         & 22                           & $4.50\times 10^{33}$  & 0.3           & -              \\
B0906-49    & 88.72                & {[}87.09; 89.39{]}                       & 112                         & 670                          & $4.90\times 10^{35}$  & 3$^{(4)}$     & SPWN           \\
J0954-5430  & 85.26                & -                                        & 171                         & -                            & $1.60\times 10^{34}$  & 0.433         & -              \\
J1301-6310  & 82.36                & -                                        & 186                         & -                            & $7.60\times 10^{33}$  & 1.458         & -              \\
B0114+58    & 82.21                & {[}46.60; 89.76{]}                       & 275                         & 154                          & $2.20\times 10^{35}$  & 1.768         & -              \\
B0941-56    & 80.76                & {[}28.39; 89.79{]}                       & 323                         & 33                           & $3.00\times 10^{33}$  & 0.406         & -              \\
J0622+3749  & 77.95                & -                                        & 208$^{(5)}$                 & -                            & $2.70\times 10^{34}$  & 1.6$^{(5)}$   & TeV Halo       \\
J0905-5127  & 74.97                & -                                        & 221                         & -                            & $2.40\times 10^{34}$  & 1.33          & -              \\
B0540+23    & 68.04                & -                                        & 253                         & -                            & $4.10\times 10^{34}$  & 1.565         & Candidate Halo \\
J0633+1746  & 64.50                & {[}13.19; 87.16{]}                       & 342                         & 152                          & $3.20\times 10^{34}$  & 0.19          & TeV Halo, SPWN \\
J1549-4848  & 54.26                & -                                        & 324                         & -                            & $2.30\times 10^{34}$  & 1.308         & -              \\
B0922-52    & 52.24                & -                                        & 335                         & -                            & $3.40\times 10^{33}$  & 0.513         & -              \\
J1846+0919  & 48.28                & -                                        & 360                         & -                            & $3.40\times 10^{34}$  & 1.53          & -              \\
J0426+4933  & 46.82                & -                                        & 371                         & -                            & $2.00\times 10^{33}$  & 1.795         & -              \\
J1741-2054  & 40.16                & {[}6.28; 82.77{]}                        & 386                         & 155                          & $9.50\times 10^{33}$  & 0.3           & SPWN           \\
B0959-54    & 38.51                & -                                        & 441                         & -                            & $6.90\times 10^{32}$  & 0.3           & -              \\
B0940-55    & 36.52                & -                                        & 464                         & -                            & $3.10\times 10^{33}$  & 0.3           & -              \\
J1849-0317  & 34.76                & -                                        & 481                         & -                            & $2.90\times 10^{33}$  & 1.212         & -              \\
J2030+4415  & 29.69                & -                                        & 555                         & -                            & $2.20\times 10^{34}$  & 0.72          & -              \\
J2017-2737  & 28.55                & -                                        & 583                         & -                            & $2.10\times 10^{34}$  & 1.549         & -              \\
J1957+5033  & 21.55                & -                                        & 838                         & -                            & $5.30\times 10^{33}$  & 1.365         & -              \\
B0727-18    & 21.22                & {[}2.97; 74.07{]}                        & 426                         & 177                          & $5.60\times 10^{33}$  & 2             & -              \\
J0835-3707  & 20.94                & -                                        & 877                         & -                            & $2.40\times 10^{33}$  & 0.555         & -              \\
J1152-6012  & 20.69                & -                                        & 893                         & -                            & $4.90\times 10^{33}$  & 1.367         & -              \\
J0212+5222  & 20.58                & -                                        & 904                         & -                            & $4.90\times 10^{33}$  & 1.558         & -              \\
J1848+0647  & 20.43                & -                                        & 916                         & -                            & $2.70\times 10^{33}$  & 1.128         & -              \\
J1530-5327  & 19.95                & -                                        & 944                         & -                            & $8.50\times 10^{33}$  & 1.123         & -              \\
B1055-52    & 17.66                & {[}1.10; 86.20{]}                        & 535                         & 21                           & $3.00\times 10^{34}$  & 0.093         & -              \\
B1742-30    & 11.36                & {[}0.71; 82.23{]}                        & 546                         & 47                           & $8.50\times 10^{33}$  & 0.2           & -              \\
B0355+54    & 9.20                 & {[}0.55; 77.19{]}                        & 564                         & 59                           & $4.50\times 10^{34}$  & 1             & SPWN           \\
B0919+06    & 0.63                 & {[}\textless{}0.001; 2.35{]}             & 497                         & 461                          & $6.80\times 10^{33}$  & 1.1           & -              \\
J0357+3205  & \textless 0.001      & {[}\textless{}0.001; \textless{}0.001{]} & 540                         & 649                          & $5.90\times 10^{33}$  & 0.835         & SWPN           \\
B1133-55    & \textless 0.001      & {[}\textless{}0.001; 0.01{]}             & 702                         & 318                          & $6.70\times 10^{33}$  & 1.518         & -              \\ \hline
\end{longtable}
\end{appendix}

\end{document}